\documentclass{eas}
\usepackage{graphicx}
\usepackage{amssymb}
\begin{document}

\title{Magnetic properties of intermediate-mass stars}

\runningtitle{G.A. Wade: Stellar magnetic properties \dots}
\author{Gregg A. Wade}
\address{Department of Physics, Royal Military College of Canada}
%
%
%
%
\begin{abstract}
Magnetic fields play an important role in producing and modifying the photospheric chemical peculiarities of intermediate-mass main sequence stars. This article discusses the basic theory and methods of measurement used to detect and characterise stellar magnetic fields, and reviews our current knowledge of selected characteristics of magnetic fields in intermediate-mass stars.
  
\end{abstract}
\maketitle
\section{Introduction}

Magnetic fields play a fundamental role in the physics of the atmospheres of a significant fraction of stars on the H-R diagram. The magnetic fields of early-type stars (mainly the intermediate-mass main sequence Ap and Bp stars) have quite different characteristics (and probably a different origin) than those of late-type stars like the sun (e.g. Mestel 2003). In early-type stars, the surface magnetic field is static on a timescale of at least many decades, and appears to be "frozen'' into a rigidly rotating atmosphere. The magnetic field is organized on a large scale, permeating the entire stellar surface, with a relatively high field strength (typically of a few hundreds up to a few tens of thousands of gauss). The presence of a magnetic field strongly influences energy and mass transport within the atmosphere, and results in the presence of strong chemical abundance nonuniformities in photospheric layers (one of our reasons for having this meeting!). Due its global nature, the magnetic field of Ap stars can be much more easily detected and studied than that of late-type stars. Therefore, Ap stars represent a primary target for spectropolarimetric observations, for the development and testing of the instruments and modelling techniques, and for evaluation of physical models.ssssssssss

\section{Detection and measurement of stellar magnetic fields}

Stellar magnetic fields are detected and characterised using the Zeeman 
effect. When an atom is immersed in an external magnetic field $B$, 
individual atomic levels (with energy $E_0$, total angular momentum $J$, and 
Land\'e factor $g$) are split into $(2J+1)$ substates, characterised by their 
magnetic quantum number $M\ (-J\leq M\leq +J)$. The energies of these atomic 
states are given by:
\begin{equation}
E(M) = E_0 gM\hbar \omega_L
\end{equation}
where $\omega_L=eB/(2m_ec)$ is the Larmor frequency (e.g. Mathys 1989). 

As a result of this splitting, each transition that generates a single spectral line 
in the stellar spectrum when no field is present leads, in the presence of a 
field, to a group of closely-spaced lines, or {\em Zeeman components}.
For transitions in LS coupling, these lines may be grouped into two different 
types with different properties. Those lines resulting from transitions in 
which $M$ does not change (i.e. $\Delta M=0$) are spread symmetrically about 
the zero-field wavelength $\lambda_0=hc/E_0$ of the line. These are called 
the $\pi$ components. Those resulting from transitions in which $\Delta 
M=\pm 1$ have wavelengths to the red $(+)$ and blue $(-)$ of the zero-field 
wavelength. These are called the $\sigma$ components. The wavelength 
separation between the centroids of the $\pi$ and $\sigma$ component groups 
can be calculated from Eq. (1), and is:
\begin{equation}
\Delta\lambda = {{eN\bar g}\over{4\pi m_e c^2}} \lambda_0^2\equiv \Delta\lambda_Z B\bar g
\end{equation}
where $\bar g$ is the effective Land\'e factor of the transition and 
$\Delta\lambda_Z$ is called the {\em Lorentz unit}. Note that the splitting is
directly proportional to the strength of the applied field $B$. For 
illustration, a 5 kG magnetic field produces a splitting of $\sim 0.1$~\AA\ 
at 5000~\AA. 


Pieter Zeeman was the first to report these splitting properties in his 
seminal paper of 1897, {\em The Effect of Magnetisation on the Nature of Light 
Emitted by a Substance}:
\begin{quote}
Sodium was strongly heated in a tube of biscuit porcelain, such as Pringsheim 
used in his interesting investigations upon the radiation of gases. The tube 
was closed at both ends by plane parallel glass plates, whose effective area 
was 1 cm. The tube was placed horizontally between the poles, at right angles 
to the lines of force. The light of an arc lamp was sent through. The 
absorption spectrum showed both D lines. The tube was continuously rotated 
round its axis to avoid temperature variations. {\em Excitation of the magnet 
caused immediate widening of the lines. It thus appears very probable that 
the period of sodium light is altered in the magnetic field.}
\flushright{(Zeeman, 1897)}
\end{quote}

At the suggestion of 
Hendrik Lorentz, Zeeman furthermore investigated the polarisation properties 
of the $\pi$ and $\sigma$ components. Zeeman found that in a magnetic field 
aligned parallel to the observer's line-of-sight (a {\em longitudinal} 
field), the $\pi$ components vanish, whereas the $\sigma$ components have 
opposite circular polarisations. On the contrary, when the magnetic field is 
aligned perpendicular to the observer's line-of-sight (a {\em transverse} 
field), the $\pi$ components are linearly polarised parallel to the field 
direction, while the $\sigma$ components are linearly polarised perpendicular 
to the field direction:

\begin{quote}
I have since found by means of a quarter-wave plate and an analyser, that 
{\em the edges of the magnetically-widened lines are really circularly 
polarised when the line of sight coincides in direction with the lines of 
force...} On the contrary, {\em if one looks at the flame in a direction at 
right angles to the lines of force, then the edges of the broadened sodium 
lines appear plane polarised}, in accordance with theory.
\flushright{(Zeeman, 1897)}
\end{quote}

In summary, Zeeman observed that application of a magnetic field leads to 
splitting of spectral lines into multiple Zeeman components, and that the 
components are polarised in particular ways corresponding to the orientation 
of the field. Thus, both the {\em intensity} and the {\em geometry} of the 
magnetic field in the line-forming region are encoded in the line profile as 
a consequence of Zeeman effect. 

In the ``weak-field limit'', when the ratio of the Zeeman splitting to the 
intrinsic width of the line $\Delta\lambda_Z/\Delta\lambda_I\ll 1$, under the 
assumption of a linear Milne-Eddington source function, one obtains the 
first-order solution for the Stokes profiles emergent from a stellar 
photosphere:
\begin{eqnarray}
I(\tau=0, \lambda) & = & B_0 + B_0\beta_0\mu [1 + \eta(\lambda)]^{-1} \\
Q(\tau=0, \lambda) & = & 0\\
U(\tau=0, \lambda) & = & 0\\
V(\tau=0, \lambda) & = & -\bar g\Delta\lambda_Z \langle B_z\rangle {dI(\lambda) \over d\lambda}
\end{eqnarray}
where $B_0$ and $\beta_0$ and the Milne-Eddington parameters, $\mu$ is the 
limb angle, $\eta(\lambda)$ is the line opacity variation in the absence of a 
magnetic field, and $\langle B_z\rangle$ is the longitudinal component of the magnetic field.

Eqn.~(2.3) will be immediately recognised as the Milne-Eddington solution for 
the emergent Stokes $I$ profile in the {\em absence of a magnetic field} - 
i.e., to first order, the Stokes $I$ profile is unaffected by the presence of 
a weak magnetic field. Furthermore, Eqs.~(2.4) and (2.5) show, to first order, 
that the line profile is not linearly polarised for weak fields. In fact, the 
only first-order influence of the magnetic field in the weak-field regime is 
on the amplitude of the Stokes $V$ profile. For any spectral line, the shape 
of Stokes $V$ in the weak-field regime is determined entirely by the shape of the corresponding Stokes 
$I$ profile, via the derivative $dI/d\lambda$. However, the amplitude of 
Stokes $V$ is proportional to the intensity of the longitudinal component of 
the magnetic field $\langle B_z\rangle$. This implies that for most stars, the most easily 
accessible Zeeman diagnostic will be line circular polarisation, and for such 
stars only the longitudinal field component can be practically measured. The 
linear polarisation Stokes parameters, which constrain the transverse 
components of the field, will be in general much weaker (this is a 
2$^{\rm nd}$ order effect), and consequently much more difficult to detect. 

In real stars, our ability to diagnose magnetic fields depends on various 
(non-magnetic) physical and spectroscopic attributes (e.g. visual magnitude, 
S/N, spectral line depth and density, rotational velocity, spectral resolving 
power of the spectrograph), as well as the magnetic field. In many situations,
the observable signal may be near the practical limits of available 
instrumentation. 


\section{Application}

\begin{figure}[t]
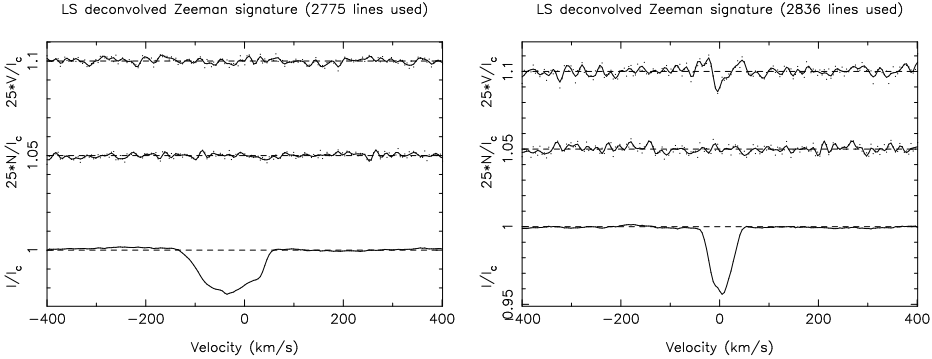

\begin{center}
\includegraphics[height=6.0cm,angle=-90]{coolfig4a.ps}\hspace{2mm}
\includegraphics[height=6.0cm,angle=-90]{coolfig2b.ps}
\caption{LSD Stokes $I$ and $V$ mean profiles for two classified cool Ap stars: {\em Left --}\ HD 182640 ($B_z=-60\pm 41$~G). {\em Right --}\ HD 96707 ($B_z=+1\pm 37$~G). Although both stars exhibit no significant detection of the longitudinal magnetic field, circular polarisation (and hence magnetic field) is clearly detected in the case of HD 96707.}
\label{fig-2}
\end{center}
\end{figure}

As discussed above, the most easily accessible polarised diagnostic of stellar magnetic fields is circular polarisation (resulting from the longitudinal Zeeman effect and indicative of the presence of a line-of-sight component of the magnetic field in the stellar photosphere (a longitudinal magnetic field)).
Typically, Zeeman circular (or linear) polarimetry is obtained {\em differentially}, by recording two
 orthogonal (circular or linear) polarisation states, separated using a $\lambda/4$ (or $\lambda/2$) retarder and a polarising beamsplitter) simultaneously
 on a CCD\footnote{In fact, most modern methods employ a rotation of the retarder to permit switching of the positions of the two beams on the CCD, allowing the
 removal of systematic detector effects with multiple exposures; see, e.g. Semel
 et al. (1993).}. The sum of the recorded fluxes gives the natural light spectrum (Stokes $I$), whereas their difference gives the net circularly (or linearly) polarised flux
 (Stokes $V$ or $Q/U$). This procedure is described in some detail by e.g. Bagnulo et al. (2005a).

Zeeman circular polarimetry has the advantage of being sensitive even under conditions of moderate or rapid stellar rotation. Using metal lines, practical sensitivity up to about 60~${\rm km}\, {\rm s}^{-1}$ can be achieved for Ap stars with strong fields (e.g. Wade et al. 2000a)\footnote{Although limited in sensitivity to moderate $v\sin i$ objects, metallic line spectropolarimetry can exploit stellar rotation to allow detection of magnetic fields {\em even when the mean longitudinal magnetic field is null} (see e.g. Shorlin et al. (2002). See also Fig. 1).}. When metallic-line spectropolarimetry is coupled with the powerful Least-Squares Deconvolution multi-line technique (LSD, Donati et al. 1997, Fig. 2), remarkably high precision can
 be attained, particularly for cooler A stars. Using intrinsically broad lines such as Balmer lines, sensitivity can furthermore be extended to several hundreds
 of ${\rm km}\, {\rm s}^{-1}$ (e.g. Bagnulo et al. 2002, see also e.g. Bohlender et al. 1987; see Fig. 2)\footnote{With the disadvantage that such techniques are not sensitive to null longitudinal fields.}. These techniques have been recently employed in studies by e.g. Shorlin et al. (2002), Leone \& Catanzaro (2004), Chadid et al. (2004) and Bagnulo et al. (2004), achieving in some cases longitudinal field precision better than 10 G. Notably, circular polarimetry of metallic lines by Babcock, Preston and collaborators, and
in Balmer lines by Landstreet, Borra and collaborators, have led the way to the
fundamental understanding of magnetism in intermediate mass stars that we enjoy
today.

\begin{figure}[t]
\centerline{\includegraphics[width=6cm]{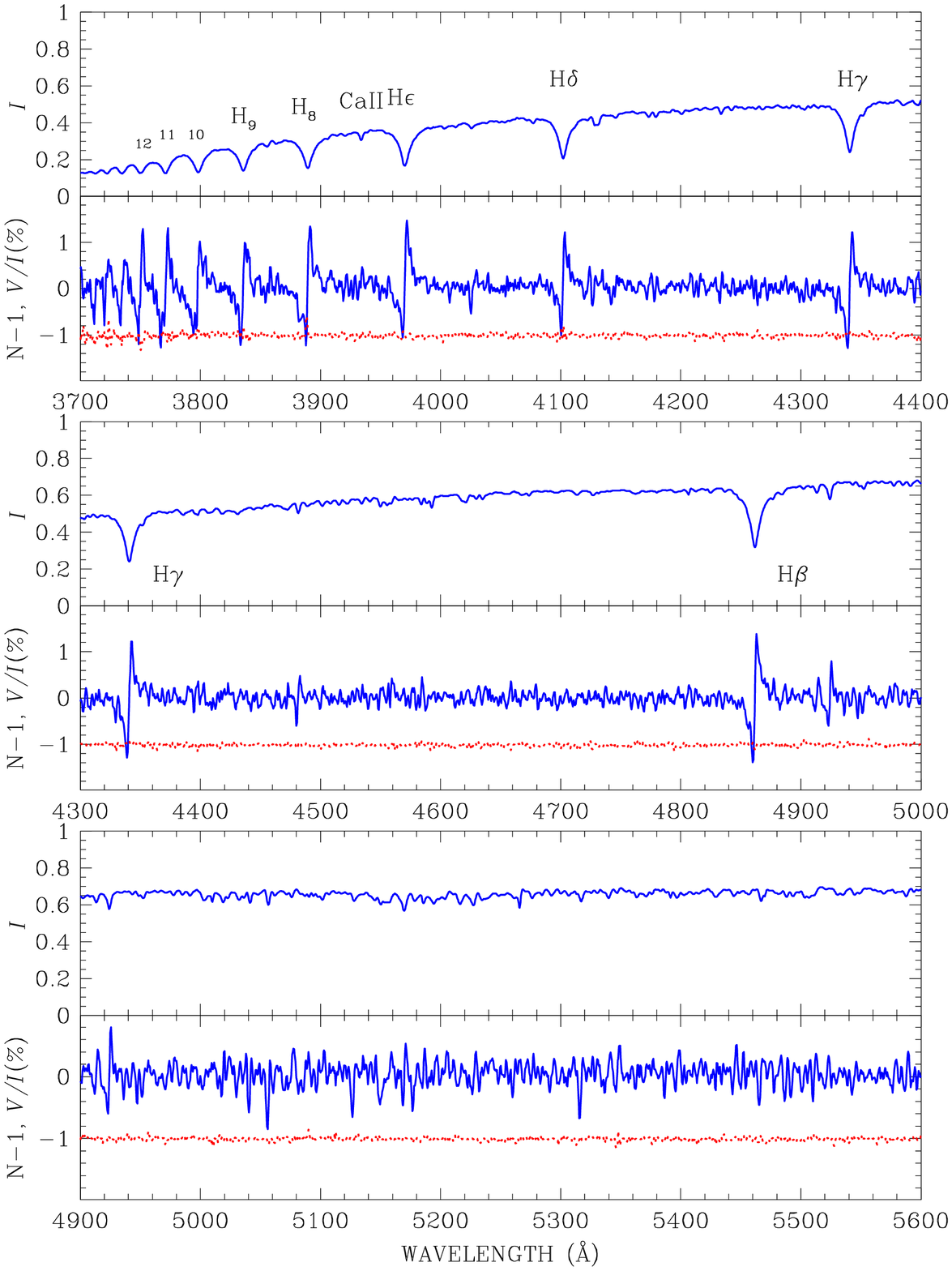} \includegraphics[width=6cm]{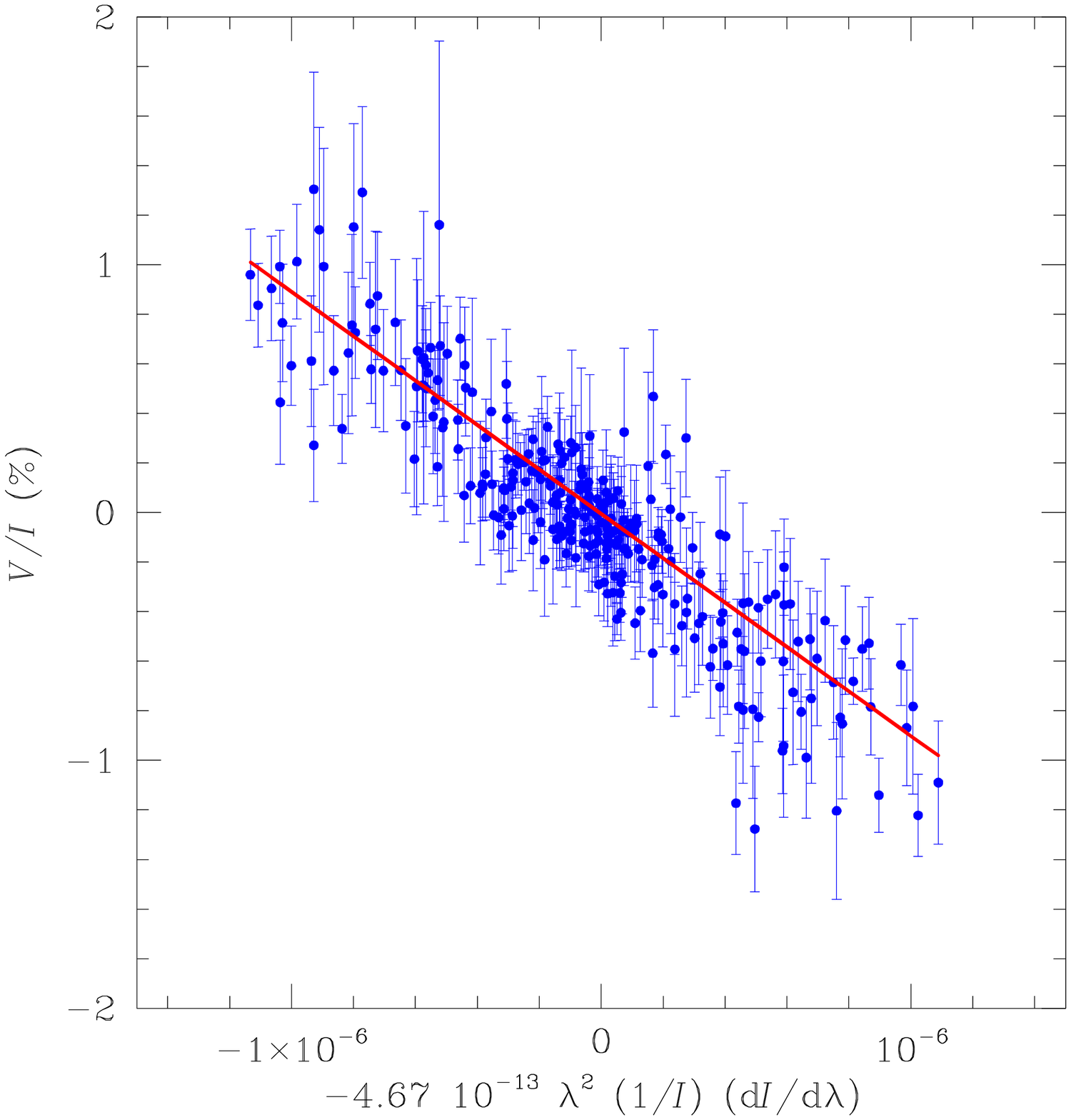}}\caption{Detection of a -9 kG longitudinal magnetic field in NGC 2244-334. {\em Left -}\ FORS1 Stokes $I$ and $V$ spectrum showing strong circular polarisation profiles associated with all Balmer series lines. {\em Right -}\ Regression providing diagnosis of the longitudinal magnetic field according to Eq. (2.6).}
\end{figure}

By observing Zeeman linear polarisation (the transverse Zeeman effect, sensitive to the plane-of-sky componenent of the magnetic field) in combination with the circular polarisation longitudinal Zeeman effect, qualitatively new constraints on the magnetic field topologies of intermediate-mass stars can be obtained. The first systematic observations of the transverse Zeeman effect in A stars were obtained by Leroy and collaborators (e.g. Leroy 1995) during the early 1990s. These
authors measured time variations of low levels of linear polarisation in the broadband light of Ap stars, produced as a result of saturation of Zeeman-broadened
 metallic absorption lines. More recently, observations of transverse Zeeman effect {\em within} spectral absorption lines has become possible (Wade et al. 2000b obtained high-resolution observations of several stars using the MuSiCoS spectropolarimeter at Pic du
Midi observatory).

\section{Magnetic properties of intermediate-mass stars}

In this section we describe the current state of understanding of selected properties of magnetic upper main sequence stars, as inferred using the methods described above.

\subsection{Configuration and structure}

Spectroscopic, polarimetric and spectropolarimetric observations of intermediate-mass stars generally imply the presence of static, globally-ordered magnetic fields with important dipole components, tilted with respect to the stellar rotational axis. The tilt of the dipole axis has been found to correlate rather strongly with stellar rotational period (Landstreet \& Mathys 2000), with the most slowly-rotating stars exhibiting nearly aligned magnetic and rotational axis, whereas more rapidly rotating stars show much larger tilt angles. 

Historically, a few clear examples of departures of the large-scale field from a (centred or decentred) dipolar configuration have been reported (e.g. Borra \& Landstreet (1980), Thompson \& Landstreet 1985, Mathys \& Lanz 1997). The particular case of the helium-strong star HD 37776 (Thompson \& Landstreet 1985) deserves special attention. This star displays a clearly double-peaked longitudinal magnetic field variation, which was argued by Thompson \& Landstreet (and later by Bohlender 1988) as the signature of a dominantly quadrupolar magnetic field.

One striking difference as compared to lower main sequence stars is the near-unity filling factor and low field intensity contrast of upper main sequence magnetic stars. Significantly, the presence of weak fields occupying an important fraction of the stellar surface (and therefore a filling factor substantially different from unity) can be ruled out using magnetically-split spectral lines (Mathys et al. 1997) and the Magnetic Doppler Imaging method (e.g. Piskunov \& Kochukhov 2002).

In recent years, various authors have used increasingly sophisticated data to demonstrate more generally the limitations of an axisymmetric dipolar configuration (Leroy et al. 1995, Bagnulo et al. 2002). Most recently, modelling of high resolution, high signal-to-noise ratio Stokes $IQUV$ line profiles using Magnetic Doppler Imaging (MDI) by Kochukhov et al. (2004, and in preparation) has yielded high-resolution maps of the surface magnetic field topologies of two Ap stars, 53 Cam and $\alpha^2$~CVn. As described in the paper by Kochukhov in this volume, although the surface magnetic field structure of $\alpha^2$~CVn displays only mild departures from a dipolar configuration (in both intensity and topology), 53 Cam shows small-scale structure in the intensity of its surface magnetic field that was totally unexpected based on low-resolution global field models. These results suggest that there exists significant diversity in the surface structure of magnetic fields of intermediate-mass stars - diversity that can only be studied using high-resolution spectropolarimetric measurements and surface imaging techniques such as MDI.

\begin{figure}
\centerline{  \includegraphics[width=7.5cm]{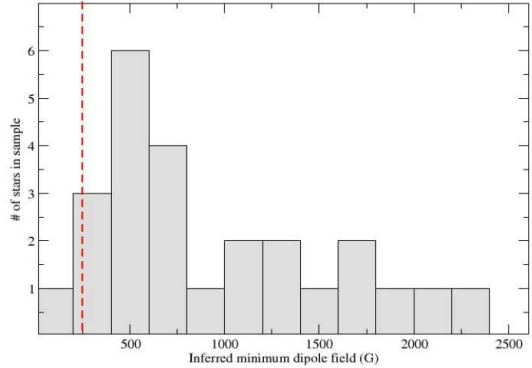}}
  \caption{Distribution of minimum surface dipole strengths of a sample of about 30 Ap stars, from the work of Auri\`ere et al. (this volume). Note the significant lack of stars with minimum surface dipole strengths below 300 G.}
\end{figure}

\begin{figure}[h]
\centerline{  \includegraphics[width=7.5cm,angle=-90]{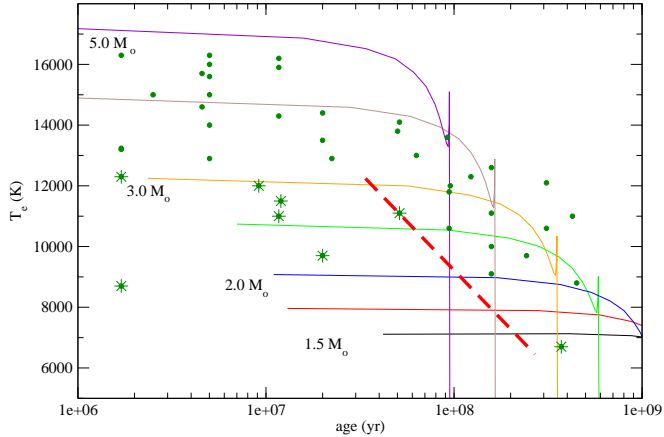}}
  \caption{Age/effective temperature HR diagram of open cluster Ap/Bp stars in which magnetic fields were detected by Bagnulo et al. (2005). Note the identification of several very young magnetic stars with masses in the range 2-3~$M_\odot$ which have completed less than 10\% of their main sequence evolution.}
\end{figure}
\subsection{Incidence}

Past searches for magnetic fields in A and B type stars have strongly suggested that a ``magnetic dichotomy'' exists on the upper main sequence: that the Ap and Bp stars (about 5\% of the main sequence stars between about F0 and B0) host magnetic fields, whereas the rest do not. A few non-Ap stars have been suggested to host magnetic fields (e.g. Lanz \& Mathys 1993), but such claims were never confirmed. More recently, based on a survey of line circular polarisation in a sample of 74 A and B-type stars, Shorlin et al. (2002) concluded that ordered magnetic fields must be very rare or completely non-existent in non-Ap/Bp intermediate-mass stars. They were able to obtain only upper limits for all non-Ap/Bp stars in their sample (no fields were detected in any of these objects), constraining the longitudinal magnetic fields of about 13 stars to under 10 G, and about twice that many to be under 20 G. Significantly, Shorlin et al. were also able to rule out the presence of complex magnetic field topologies, similar to those of active late-type stars, based on their velocity-resolved Stokes $V$ profiles. This study provides very strong evidence that the majority of A and B stars do not display any evidence of magnetic fields in their photospheres. On the other hand, recent investigations by Auri\`ere et al. (this volume), who observed a sample of bright, spectroscopically-classified Ap/Bp stars in which magnetic fields have generally never been detected, demonstrate that all Ap stars, when observed with sufficient precision, show evidence for ordered magnetic fields with surface intensities of at least $\sim 300$~G (approximately the equipartition field in the photosphere around $\tau=1.0$; see Fig. 3). These two studies strongly support the existence of this ``magnetic dichotomy'' amongst intermediate-mass main sequence stars.

\subsection{Origin/appearance/evolution}

Recent work by Pohnl et al. (2003, 2005), and Bagnulo et al. (2003, 2004, 2005a, and in preparation) exploiting open clusters has firmly established the existence of magnetic A and B type stars of a variety of masses (from about $2~M_\odot$ to over $5~M_\odot$) at the very earliest stages of their main sequence evolution (having completed less than 10\% of their main sequence evolution; see Fig. 4). Drouin (2005), Wade et al. (2005) and Catala et al. (in preparation) have furthermore succeeded in extending work of Donati et al. (1997), detecting and characterising magnetic fields in several pre-main sequence stars of intermediate mass (see Fig. 5). Based on a large sample of stars and detailed investigation of individual objects, these authors conclude that the magnetic properties of the pre-main sequence A and B stars (statistical incidence of magnetism, surface intensity of the fields, and surface structure of the fields) are quantitatively similar to those of their main sequence descendents. In the youngest magnetic pre-main sequence stars, they find no evidence for photospheric chemical peculiarity or non-uniformity. On the other hand, in the most evolved star studied (HD 72106A), clear evidence for strong chemical peculiarity and abundance spots is presented. They conclude that their results are consistent with the appearance of magnetic fields in intermediate-mass stars prior to arrival on the main sequence (in fact during shell deuterium burning), a result which may significantly constrain field origin theories.

Bagnulo et al. (in preparation; see Fig. 4), who have obtained measurements of the longitudinal magnetic fields of open cluster member Ap stars, have furthermore been able to examine the rough statistical evolution of field intensity along the main sequence.  Based on cumulative
histograms of field strength for the mass ranges 2-3 $M_\odot$ and 3-4~$M_\odot$, they find marginally significant differences in the field
strength distributions, in the sense that older stars in both mass ranges
tend to have smaller fields than younger stars (age~$< 5\times 10^7$~yr) stars. This result is qualitatively consistent with the hypothesis of conservation of magnetic flux as the star expands during its main sequence evolution. They also conclude that young stars can host very strong 
fields (as demonstrated by the strong magnetic fields of the very young stars HD 66318 (Bagnulo et al. 2003) and NGC-2244-334 (Bagnulo et al. 2004).

\begin{figure}
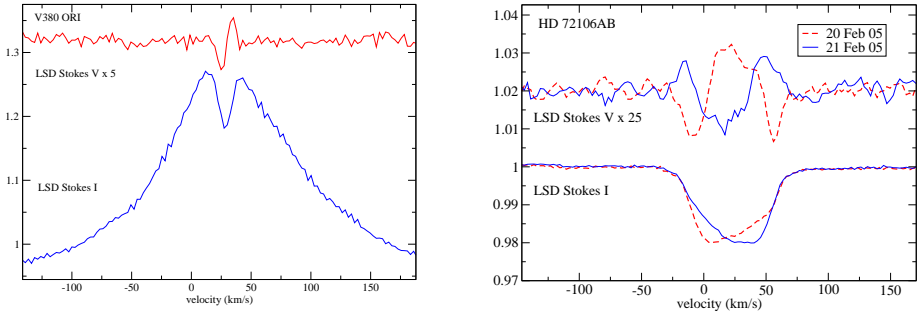

  \includegraphics[width=6cm]{v380ori-lsd.eps}\hspace{0.5cm}\includegraphics[width=6cm]{hd72106-lsd.eps}
  \caption{Zeeman circular polarisation detected in LSD profiles of the pre-main sequence stars of intermediate mass V380 Ori and HD 72106A (Wade et al. (2005). }
\end{figure}

\section{Conclusion}

Using the same basic physical principles demonstrated by Zeeman (1897), modern spectropolarimetric methods are being combined with sophisticated data analysis techniques to clarify the magnetic properties of intermediate mass stars. Some of the most important results of recent investigations are:
\begin{enumerate}
\setlength{\itemsep}{-4pt}
\item When viewed at high resolution using Magnetic Doppler Imaging, the surface magnetic field topologies of A and B stars show significant diversity (see Kochukhov, this volume).
\item A ``magnetic dichotomy'' is confirmed to exist for main sequence A and B star (see Shorlin et al. 2002, Auri\`ere et al., this volume).
\item Magnetic Ap/Bp stars of low mass (2-3~$M_\odot$) have been firmly established at very early phases of their main sequence evolution (see Pohnl et al. 2003, 2005 Bagnulo et al. 2005a, and in preparation).
\item The first magnetic pre-main sequence progenitors of magnetic Ap/Bp stars appear to have been identified (see Drouin 2005, Wade et al. 2005). 
\end{enumerate}


%


\end{document}